\documentclass{aa}
\usepackage{graphics}
\begin{document}

\thesaurus{
       (11.06.2;
            11.19.2;
       11.19.6)}

\title{Some statistical properties of spiral galaxies}
\author{J. Ma\inst{1,2,3,4}, J.L. Zhao\inst{5,1,2,4}, C.G. Shu\inst{1,2,4} and
Q.H. Peng\inst{6}}

\institute{Shanghai Astronomical Observatory, Chinese Academy of Scienes,
Shanghai, 200030, P.R. China
\and National Astronomical Observatories, Chinese Academy of Scienes, 100012, P.R. China
\and Beijing Astrophysics Center (BAC), Beijing, 100871, P.R. China
\and Joint Lab of Optical Astronomy, Chinese Academy of Scienes, P.R. China
\and CCAST (World Lab., Beijing, 100080, P.R. China)
\and Department of Astronomy, Nanjing University, Nanjing, 210093, P.R. China}

\offprints{majun@center.shao.ac.cn}

\date{Received 4 January 1999 / Accepted 26 August 1999}
\titlerunning{Some statistical properties of spiral galaxies}
\authorrunning{Ma et al.}
\maketitle

\begin{abstract}

This paper presents
some statistical correlations of
72 northern spiral galaxies. The results show that
early-type spirals that are brighter, and
thicker, and the axis ratios ($H_{z}/H_r$) of the disk tend
to be smaller along the Hubble sequence.
We also find that
$H_{z}/H_r$ correlates strongly with the galaxy's color,
and early-type spirals have larger values of $H_{z}/H_r$.
The inclinations obtained by fitting the pattern of a spiral
structure with a logarithmic spiral form are nearly the
same as those obtained
by using the formulas of Aaronson et al. (1980). Finally,
the mean measured
pitch angles for the different Hubble sequences in the
Third Catalogue of Bright Galaxies by de Vaucouleurs et al.
(1991) are derived.


\end{abstract}

\section{Introduction}

The inclination of a spiral galaxy ( i.e., the angle between the galactic plane and
the tangent plane of the celestial sphere) is not only
an important parameter, but also difficult to determine.
A spiral
galaxy consists of a thin disk, a bulge and spiral arms that
are thought to be situated in the disk. If we assume that
the thickness of the spiral plane is rather
negligible in comparision to
its extension, and that when a spiral galaxy
is inclined moderately to the plane of sky, the thickness
of the nucleus can be omitted, the inclination $(\gamma)$ can be
obtained by:

\begin{equation}
\gamma=\arccos(\frac{d}{D}),
\end{equation}
where $D$ and $d$ are the apparent major and minor isophotal
diameters respectively. When a spiral galaxy is seen edge-on, it is
not possible to consider the thickness of the nuclear part as
negligible.
Thus, Eq. (1) cannot be
used to calculate the inclination. The reason for this is that
the apparent minor isophotal-diameter consists of two parts.
One is attributed by the disk and, another by the bulge, the latter of which will
decrease the real value of the inclination.

Considering that the disk is not infinitely thin,
Aaronson et al. (1980) corrected Eq. (1) by

\begin{equation}
\gamma=\arccos\sqrt{1.042(\frac{d}{D})^{2}-0.042}+3^{\circ}.
\end{equation}
The constant of $3^{\circ}$ is added in accordance with an empirical
recipe. A more elaborate specification
of the axial ratio for an edge-on system that depends on the Hubble
type could be justified. The thinnest galaxies are Sc spirals',
earlier types have larger bulges. Giovanelli et al. (1997) provided
an example to justify why they assumed the axial ratio of Sc galaxies to be
0.13. A smaller value of the axial ratio for an
edge-on system results in smaller derived inclinations, where the spirals
are more face-on. Besides,
if the values $D$ and $d$ are approximations due to errors, the inclination obtained
by Eq. (2) is not an exact value. Ma et al. (1997, 1998) proposed
a method to determine the inclination of a spiral galaxy by fitting
a spiral arm with a logarithmic spiral form with constant
pitch angle. They obtained the inclinations of 72 northern spiral galaxies.

The question of the mathematical form of spiral arms was recognized at the
beginning of this century (von der Pahlen, 1911; Groot, 1925). Then,
Danver (1942), Kennicutt (1981) and Kennicutt \& Hodge (1982)
systematically studied the shapes of
spiral arms. Using the method of the least squares and as many points as possible
situated on the spiral
arm in question, Danver (1942) studied a sample of 98 nearby spirals
by drawing the projected images on white paper
and then, by
copying it on the paper to be used for the measurement thanks to
transparent illumination.
Kennicutt (1981) measured the shapes of spiral arms in 113 nearby Sa-Sc galaxies
by disposing directly of photographic enlargement
and using an iterative procedure
to correct for inclination effects. He gave an initial estimate of the
inclination and pitch angle to orient the spiral to a face-on geometry, and then used
any residual sinusoidal deviations in the arm shapes to make small corrections to the
derived orientation. Using the IRAF software,
Ma et al. (1997, 1998) fitted the shapes of
spiral arms on the images, so that they could show
clearly whether the fitting was good or not.
The DISPLAY program of IRAF software can enlarge the image and change its
grey scale to minimize any personal prejudice about the regularity and prominence of
arms. But we must emphasize that the DISPLAY program of IRAF has many
variables, so the
results are not always objective. In our program, we modify z1 (minimum greylevel to
be displayed) and z2 (maximum greylevel to be displayed) in the DISPLAY program
in order to display the images clearly.
In the procedure of fitting,
we emphasize the global spiral structure, where, except for the small-scale
distortions, the arms can be represented by the logarithmic spiral forms.

There has been much interest concerning the separation of disk and
bulge components in the observed surface brightness distribution of
spiral galaxies. de Vaucouleurs (1958), for instance, established
an isophotic map of M~31 in blue light by means of direct photoelectric
scans, spaced at 10$'$ intervals in declination from +39$^{\circ}$31$'$ to
42$^{\circ}$30$'$. From photoelectric photometry, he determined
that the thickness of the flat component is about 0.8 kpc.
By assuming that a galaxy has an infinitesimally
thin disk, Freeman (1970) and Sandage et al. (1970) collected and
studied the radial distribution of the surface brightness $I(r)$ for
thirty-six S0 and spiral galaxies, and showed that $I(r)$ distribution
for these galaxies can be presented by two main components: an
inner spheroidal component which follows the law of
\begin{equation}
\log I(r) \propto r^{1/4}
\end{equation}
and an outer exponential component (disk), with
\begin{equation}
\log I(r)=I_0e^{-r/h_r},
\end{equation}
where $h_r$ is defined as a radial scale length.
Van der Kruit and Searle (1981a) proposed a model for
the light distribution in the disks of edge-on spiral galaxies,
assuming that a galaxy has a locally isothermal, self-gravitating and
truncated exponential
disk. This model has the feature of being isothermal in $z$ at all
radii with a scale parameter $z_0$ and has an exponential
dependence of surface brightness upon $r$ with a scale length $h_r$.
The space-luminosity of this model can be described by
\begin{equation}
L(r, z)=L_0e^{-r/h_r}{\rm{sech}}^{2}(z/z_0).
\end{equation}
With this model, van der Kruit \& Searle (1981a, 1981b, 1982a, 1982b)
determined $h_r$ and $z_0$ for seven disk-dominated and one
spheroid-dominated spiral galaxies by using
three-color surface photometry.
Peng et al (1979) investigated three-dimensional disk galaxies, based
on the fundamental assumption by Parenago that the density distribution along
$z$-direction for a finite thickness disk is
\begin{equation}
\rho(r, \phi, z)=\frac{1}{H_z}\sigma(r, \phi)e^{-|z|/h},
\end{equation}
where $h$ is defined as an exponential scale height, $H_{z}$
is defined as a thickness
of disk and equals $2h$, and $\sigma(r, \phi)$ is
the surface density. By solving Poission's equation
for a logarithmic density perturbation, Peng et al. (1979)
obtained a criterion for density waves to appear, which is
\begin{equation}
r>r_0=\frac{H_z\sqrt{m^2+\Lambda^2}}{2},
\end{equation}
where $(r_0, \phi_0)$ is the polar coordinate of the
starting point from which arms of a galaxy stretch outward on
the galactic plane, and $m$ is the number of the arms in a
spiral galaxy. Based on this criterion, Peng (1988) proposed
a method for estimating the thickness of a non-edge-on spiral, and
derived the thicknesses of four galaxies.

Guthrie (1992) derived the axial ratios $R$ of disc components for
262 edge-on spiral galaxies on print copies of the blue Palomar
Sky Survey plates by using a microscope fitted with a micrometer
eyepiece. He then analyzed the distribution of isophotal axial ratios
for 888 diameter-limited normal Sa-Sc galaxies to give information
on the true axial ratios $R_0$, and at last presented the mean value
of $\log R_0$ is $0.95\pm0.03$.

Ma et al. (1997, 1998) derived the thicknesses of 72 spirals by
using Peng's proposal (Peng, 1988) and presented some statistical correlations
between thickness or flatness and other parameters.

The value of $R_0$ derived by Guthrie (1992) is based on observational
work and, should be much more reliable. So, it is important to compare
Ma et al.'s results (1997, 1998) to Guthrie's (1992).
In Ma et al. (1997, 1998),
the flatnesses for 72
galactic disks ($H_{z}/D_0$)
\footnote{$D_0$, which is measured at or reduced down to the
surface brightness level
$\mu_{B}=25.0 B$ magnitudes per square arcsecond,
and corrected to the ``face-on'' ($\gamma = 0^{\circ}$). For the Galactic
extinction, but not for redshift, $D_0$ is from the
Third Catalogue of Bright Galaxies by de Vaucouleurs et al.
(hereafter RC3).}
were given, and the mean value
is $0.033 \pm 0.002$. Suppose that the value of ratio of radial scale length
($h_r$) over exponential scale height ($h$) for an average exponential
galactic disk is equal to $R_0$ from Guthrie (1992), which is 9.
From Freeman (1970), we can derive $D_0/(2h_r)\approx 5$. Thus, we
obtain $H_{z}/D_0\approx 0.023$, which is in relative agreement with
the mean value of Ma et al. (1997, 1998). At the same time, we calculate
the values of $\overline{\log R_0}$ for spirals of various types T
and list them in Table 1. T, from the RC3, is morphological types,
and $\sigma$ is the dispersion.
From this table and Table 3 of
Guthrie (1992), we can see that our results are in agreement with Guthrie's
(1992). Except for the data concerning Scd galaxies, our results
also agree with de Grijs'
(1998).

\begin{table}
\caption{Values of $\overline{\log R_0}$ for spirals of various types T.}
\begin{tabular} {c|ccc}
\hline
T & $\overline{\log R_0}$ & $\sigma$ \\
\hline
2 & $0.65\pm 0.04$ & 0.09\\
3 & $0.79\pm 0.06$ & 0.21\\
4 & $0.79\pm 0.04$ & 0.21\\
5 & $0.90\pm 0.04$ & 0.18\\
6 & $1.20\pm 0.15$ & 0.33\\
\hline
\end{tabular}
\end{table}

The structure of this paper is as following: in Sect. 2, we outline
the principles of obtaining an inclination
and a pitch angle;
Sect. 3 presents some statistical properties; and
conclusions will be shown in Sect. 4.

\section{Principles of obtaining an inclination and a pitch angle}

As we know, when the line of intersection (namely,
the major axis of the image) between
the galactic plane and tangent plane
is taken as the polar axis, it is easily proved
that

\begin{equation}
r=\rho\sqrt{1+\tan^2 \gamma\sin^2\theta}
\end{equation}
and
\begin{equation}
\tan\phi=\frac{\tan\theta}{\cos\gamma},
\end{equation}
where $r$ and $\phi$ are the polar co-ordinates in the spiral plane and
$\rho$, $\theta$ are the corresponding co-ordinates in the tangent-plane, and
$\gamma$ is the inclination. If it is possible to
represent arms by equiangular
spirals as

\begin{equation}
r=r_0e^{\lambda(\phi-\phi_0)}
\end{equation}
and
\begin{equation}
\mu=\arctan\lambda,
\end{equation}
where $r$ and $\phi$ are the polar co-ordinates on the spiral arm in the spiral plane,
and $\mu$, defined as a pitch angle, is the angle between
the tangent to the arm and the
circle with the constant $r$.
The mathematical form of Eq. (10) in the tangent plane
of the celestial sphere is

\begin{equation}
\rho(\theta, \gamma)=\rho_{0}\frac{f(\theta_0,\gamma)}{f(\theta,\gamma)}e^{{\lambda}{B(\theta, \gamma)}},
\end{equation}
here
\begin{equation}
f(\theta,\gamma)=\sqrt{\sin^{2}\theta+\cos^{2}\theta\cos^{2}\gamma}
\end{equation}
and
\begin{equation}
B(\theta, \gamma)=g(\theta,\gamma)-g(\theta_{0},\gamma),
\end{equation}
where

\noindent{$g(\theta,\gamma)=$}

\begin{equation}
\left\{ \begin{array}{ll}
\arctan(\tan\theta/\cos\gamma), & k\pi-\frac{\pi}{2}<\theta<k\pi+\frac{\pi}{2} \\
\pi+\arctan(\tan\theta/\cos\gamma), & k\pi+\frac{\pi}{2}<\theta<k\pi+\frac{3\pi}{2},
\end{array}
\right.
\end{equation}
where $k$ is an integer.
Supposing that ($\rho_i$, $\theta_i$) are the measured co-ordinates of points
of the arms, and making use of the least squares method, we set

\begin{equation}
\sum_{i=1}^{n}[\rho_{i}-\rho(\theta_{i}, \gamma)]=minimum.
\end{equation}
By direct differentiation with respect to $\lambda$, we obtain the equation for
determining the parameter $\lambda$ of

\begin{equation}
\sum_{i=1}^{n}B(\theta_{i}, \gamma)\rho(\theta_{i}, \gamma)[\rho_{i}-\rho(\theta_{i}, \gamma)]=0.
\end{equation}
This is a transcendental equation. Using the least squares method, we
can obtain $\gamma$ and $\mu$. Details have been shown in Ma et al. (1997, 1998).

\section{Statistical properties}

\subsection{Sample}

Our statistical sample contains 72 northern spiral galaxies for which the
values of disk thickness and inclination and pitch angle of individual
spiral arms are from Ma et al. (1997, 1998). Except for M~31,
the images,
which have distinguishable spiral arms, are from the
Digitized Palomar Sky Survey. The image of M~31 is
digitized from POSSII (field No. 295)
via PDS measurement at Purple Mountain Observatory of China.
Most galaxies in this sample have the grand design structure.
All the galaxies have the total color indexes ($(B-V)^0_T$), which
are corrected for
the differential Galactic and internal extinction to ``face-on'', and for redshift
between $B$ and $V$ bands. The mean numerical
Hubble stage indexes (T) of the galaxies, which are quoted in the RC3, are
from 2 to 6, and 2, 3, 4, 5 and 6 correspond to Sab, Sb, Sbc, Sc and
Scd respectively. This is an inclination-limited sample, i.e., the
values of $\log(D_{25}/d_{25}$) for these galaxies are smaller than the
limited-value (0.76), where $D_{25}$ and $d_{25}$, taken from
the RC3, are the apparent major and minor isophotal diameters
measured at or reduced down to the surface brightness level
$\mu_{B}=25.0 B$
magnitudes per square arcsecond.
Clearly, this is not a diameter-limited sample.

\subsection{Thickness vs absolute magnitude}

Ma et al. (1998) have presented some
statistical correlation between thickness or flatness and
other parameters. In this section, we will investigate some
other statistical correlations.

Fig.~1 plots the correlation between the thickness and the total (``
face-on'')
absolute magnitude in the B band.
The latter is derived by using the formula

\begin{equation}
M^{o}_{B}=B^{o}_{T}+5-5\log d,
\end{equation}
here $B^{o}_{T}$ is the total ``face-on'' magnitude
from the RC3,
corrected for the Galactic and internal extinction,
as well as for redshift, d the distance, in units of pc, from
Ma et al. (1997, 1998). One can find that there is
a significant correlation between thickness and B-band
absolute magnitude, i.e.,
thicker
galaxies are brighter.

Based on the careful studies of Binggeli et al. (1985) about
the Virgo cluster, Binggeli et al. (1988) have thoroughly
reviewed what is currently known about the luminosity
function $\Phi(L)$. They have derived the luminosity function
$\Phi(L, T)$ for each morphological type separately in
the Virgo cluster. Furthermore, according to the
detailed analysis of the member galaxies for the Virgo cluster
(Zhao \& Shao, 1994; Shao \& Zhao, 1996),
Shu et al. (1995) also investigated its
LF for individual morphological types.
Combining with the luminosity function
$\Phi(L, T)$ of the Virgo cluster (Binggeli et al.,
1988; Shu, et al., 1995), Fig.~1 automatically implies that early-type
spirals, which are on average brighter than late-type
ones, are also thicker.

\begin{figure}
\resizebox{\hsize}{!}{\rotatebox{-90}{\includegraphics{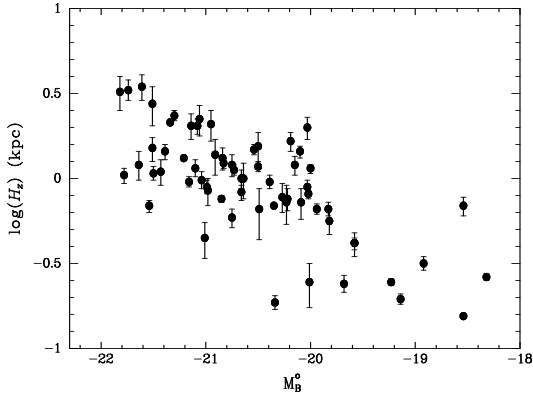}}}
\caption{Thickness of spiral galaxy plotted against the
total (``face-on'') absolute magnitude in the B-band.}
\label{FigVibStab}
\end{figure}




\subsection{Dependence of $H_{z}/H_r$ on morphological type, luminosity
and color}

In this section, we will present some statistical correlations
between $H_{z}/H_r$ and morphological type, luminosity and color for
spiral galaxies.

\begin{figure}
\resizebox{\hsize}{!}{\rotatebox{-90}{\includegraphics{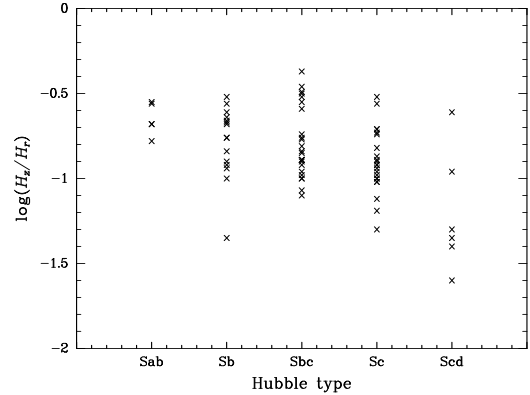}}}
\caption{$H_{z}/H_r$ of spiral galaxy plotted against the Hubble type.}
\label{FigVibStab}
\end{figure}

Ma et al. (1997, 1998) derived the flatnesses ($H_{z}/D_0$) for 72 spiral
galaxies. When we take $D_0/H_r\approx 5$ ($H_r$ is defined
as $2h_r$), we can derive the values of $H_{z}/H_r$ for these spirals.
Fig.~2 shows $H_{z}/H_r$ as a function of different Hubble types.
It can be seen that
$H_{z}/H_r$ of spiral galaxies decreases smoothly an average along
the Hubble type, but the dispersion in $H_{z}/H_r$ among galaxies of the
same Hubble sequence is very large. This, perhaps, reflects that
the intrinsic flattening of spiral disks is smaller for
later type galaxies. de Grijs (1998) also shows that galaxies
become systematically thinner when going from type S0 to Sc.

\begin{figure}
\resizebox{\hsize}{!}{\rotatebox{-90}{\includegraphics{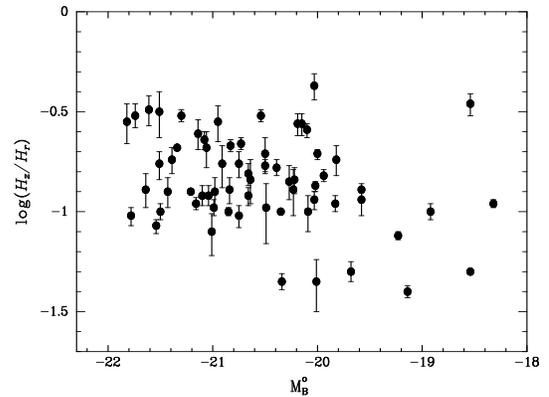}}}
\caption{$H_{z}/H_r$ of spiral galaxy plotted against magnitude in the B-band.}
\label{FigVibStab}
\end{figure}

Fig.~3 plots $H_{z}/H_r$ as a function of
total (``face-on'')
absolute magnitude in the B band.
In
Fig.~3, we can see a clear trend reflecting an
increase in the values of $H_{z}/H_r$ as the
galaxies become brighter for all but a few galaxies.
Combining this data with the luminosity function
$\Phi(L, T)$ of the Virgo cluster (Binggeli et al.,
1988; Shu et al., 1995), we find that early-type
spirals, which are brighter on average than late-type
ones, have larger values of $H_{z}/H_r$.

\begin{figure}
\resizebox{\hsize}{!}{\rotatebox{-90}{\includegraphics{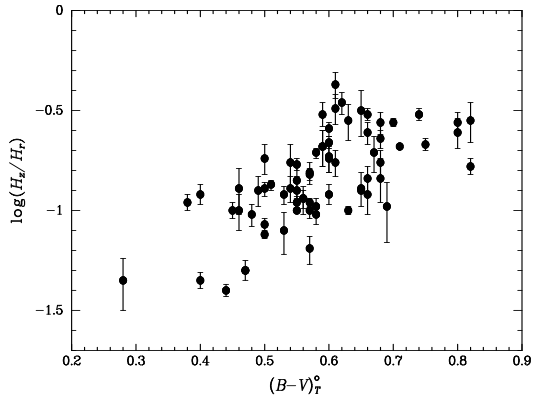}}}
\caption{$H_{z}/H_r$ of spiral galaxies plotted versus color.}
\label{FigVibStab}
\end{figure}

Fig.~4 plots $H_{z}/H_r$ as a function of
total galactic color index ($(B-V)^0_T$) in the RC3 and
shows a strong
trend which suggests that a bluer galaxy has a smaller value of $H_{z}/H_r$.
Roberts \& Haynes (1994) studied the physical parameters
along the Hubble sequence systematically by making use of two primary catalogues, the
RC3 and a private catalogue maintained by R. Giovanclli \&
M. Haynes. He presented the well-established trend between
morphology and mean color which reveals that the E and S0 galaxies are clearly
redder than spirals and that late-type spirals are bluer
on average than early-type ones.
So, Fig.~4 also implies that later-type galaxies have
smaller values of $H_{z}/H_r$.

\subsection{Statistical property of inclination}

\begin{figure}
\resizebox{\hsize}{!}{\rotatebox{-90}{\includegraphics{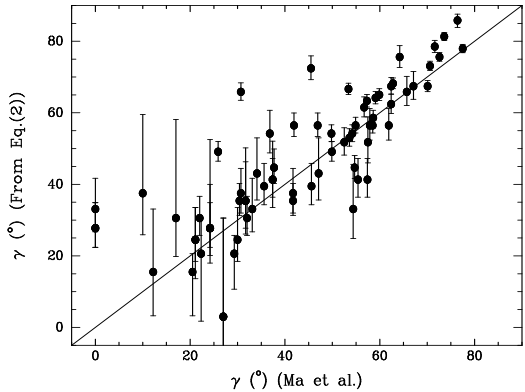}}}
\caption{Comparison of inclinations derived from Eq. (2)
and Ma et al. (1997, 1998).}
\label{FigVibStab}
\end{figure}

Tully \& Fisher (1977) showed that if a spiral structure
is well defined, the opening of the spiral structure could
be used to define the inclination of the disk. It has been shown
by Danver (1942), Kennicutt (1981) and Kennicutt \& Hodge (1982) that
the spiral arm can be represented by a logarithmic spiral form
with a constant pitch angle. Grosb$\phi$l (1985) studied 605
galaxies with inclination angles inferior to $56^{\circ}$, and
estimated the position and inclination angles of these
galaxies based on the bisymmetric intensity distribution
in their outer parts applying a one dimensional Fourier
transform method.
Ma et al. (1997, 1998) obtained the inclinations of 72 northern
spiral galaxies by fitting the arms on the images.
The RC3 lists the values of $D_{25}/d_{25}$ and their errors, $D_{25}$
and $d_{25}$ are the apparent major and minor isophotal diameters
measured at or reduced down to the surface brightness level
$\mu_{B}=25.0$ B
magnitudes per square
arcsecond. We calculate the inclinations by use of Eq. (2).
Fig.~5 presents the correlation between them and those obtained by
Ma et al. (1997, 1998). This figure shows that,
when we take into account the
errors in the inclination due to errors in $D_{25}/d_{25}$,
and although there is some scatter, Ma et al.'s result (1997, 1998) is consistent
with what is derived from Eq. (2).

\subsection{Mean value of the pitch angle along the Hubble sequences in the RC3}

Hubble (1926, 1936) introduced an early scheme to classify galaxies.
Its concepts are
still in use, a sequence starting from elliptical to spiral galaxies, and
including lenticular.
This scheme has been extended by some astronomers
(Holmberg, 1958; de Vaucouleurs, 1956, 1959; Morgan, 1958, 1959;
van den Bergh, 1960a, b, 1976; Sandage, 1961;
Sandage \& Tammann, 1981, 1987; Sandage \& Bedke, 1993)
over the years, who tried to employ multiple classification criteria.
The two main systems commonly used are
derived from Hubble's original classification criteria. One is the Hubble
system as explained in detail by Sandage (1961), Sandage \&
Tammann (1981, 1987) and Sandage \& Bedke (1993), and another, developed
by de Vaucouleurs (1956, 1959), adds more detailed descriptions to the
notation and
makes a division and extension of the Sc and SBc families by
introducing the Scd, Sd, Sdm, Sm and Im subdivisions.

Galaxy morphological classification is still mainly done visually by
dedicated astronomers, based on the Hubble's original scheme and
its modification. It is possible for each observer to give slightly
different weights to various criteria, although the criteria for
classification are accepted generally by them. Lahav et al. (1995)
and Naim et al. (1995) investigated the consistency of visual morphological
classifications of galaxies from six independent observers. They found
that individual observers agree with one another with combined rms
dispersions of between 1.3 and 2.3 type units, typically about 1.8 units
of the revised Hubble numerical type index T, although there are
some cases where the range of types given to it was superior to 4 units
in width.

The tightness of the spiral arm, in addition to the degree of resolution
in the arm and
the relative size of the unresolved nuclear region, are the fundamental
criteria in
Hubble's (1926, 1936) classification of spiral galaxies. In order to better
understand the nature and origin of the Hubble sequence and evaluate the
difference of classification, Kennicutt (1981) compared the arm pitch angles as
measured by him with the Hubble type as determined by Sandage \& Tammann
(1981, hereafter ST) and Yerkes class by Morgan (1958, 1959).
From Figs. 7 and 8 in Kennicutt (1981), we can see that, although
the ST's classification is
based almost solely upon disk resolution and the Morgan's is based entirely on the central
concentration of the galaxies,
the trends between the arm pitch angles and the two types are almost
the same. In the classifications by de Vaucouleurs et al.
(1976, RC2; 1991, RC3),  all these three criteria are considered.

A most important feature of spiral galaxies is that the stellar content
between the spiral arms and the spheroidal component is very different, and
the central part is redder than the arm region. The Yerkes system
by Morgan (1958, 1959) contains the information on the state of stellar
evolution in the central region of galaxies and it complements the information
in the Hubble sequence by de Vaucouleurs (1956, 1959), which, in cases
of conflicting criteria, emphasizes more strongly the strength of
population in the arm.
There are other sets of classifications
in which
disk-to-bulge is the primary discriminant (van den Bergh, 1976) or
the ratio of disk-to-bulge
to spiral arm morphology is primarily considered (Dressler, 1980;
Wilkerson, 1980). Among these three criteria, perhaps, spiral arm morphology may be
distance independent.

Until now, there are only a few papers dealing with pitch angles. So,
the mean value of pitch angle for the different Hubble types, which is
an important parameter in the decision of whether the WKB approximation can be
satisfied (Binney \& Tremaine, 1987), has not been presented.
As we know, the WKB approximation
is an indispensable tool for understanding the origin and
evolution of a density wave in spiral galaxies.

\begin{table}
\caption{Mean values of the pitch angles $\mu$
for different Hubble types.}
\begin{tabular} {c|cccc}
\hline
Hubble type & $\bar{\mu}$ & $\sigma$ & N\\
\hline
S0/a & $6.25\pm1.08$ & 2.86 & 8 \\
Sa & $4.60\pm1.40$ & 2.80 & 5 \\
Sab & $9.36\pm1.03$ & 3.71 & 14 \\
Sb & $11.82\pm0.81$ & 4.35 & 30 \\
Sbc & $15.16\pm0.76$ & 5.36 & 51 \\
Sc & $15.50\pm0.62$ & 4.25 & 48 \\
Scd & $18.92\pm1.19$ & 5.34 & 21 \\
Sd & $25.00\pm0.00$ & 0.00 & 2 \\
\hline
\end{tabular}
\end{table}

\begin{figure}
\resizebox{\hsize}{!}{\rotatebox{-90}{\includegraphics{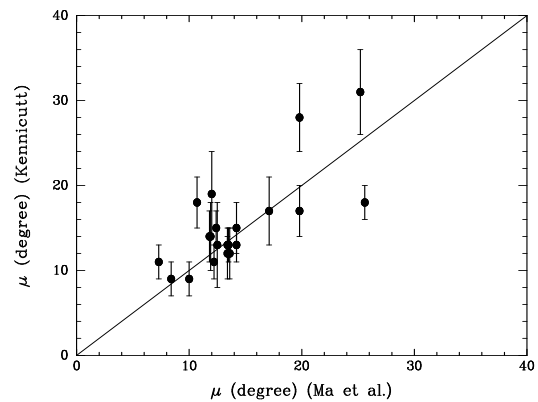}}}
\caption{Comparison of pitch angles derived by Kennicutt (1981) and Ma et al.
(1997, 1998),
for galaxies in common.}
\label{FigVibStab}
\end{figure}

Before deriving the mean value of pitch angle for the different types in
the RC3, it is useful to compare the pitch angles of the common galaxies
derived by Kennicutt (1981) and Ma et al. (1998). 22 of the spirals
measured by Ma et al. (1998) were previously done by Kennicutt (1981),
and the average pitch angles are compared in Fig.~6.
Except for 5 galaxies, the results are consistent between
Kennicutt (1981) and Ma et al. (1998). Some of the discrepancies can
be attributed to our measured error, and the others may be expected
from the galaxies themselves; for example, when a spiral galaxy has two
asymmetric arms: different authors (Kennicutt, 1981; Ma et al.
1998) did not measure the same arm in the same galaxy.

Table 2 presents
the mean values of pitch angle ($\mu$) from the ones measured by
Kennicutt (1981) and Ma et al. (1997, 1998) for the different
Hubble sequences in the RC3, where $\sigma$ is
the dispersion and
N the number of galaxies.
The important result from Table 2 is that the mean pitch angles
of Sa-Sc are not larger than $15.5^{\circ}$, so that cotan$\mu\geq3.6$.
Thus the WKB approximation is satisfied at the mean pitch
angles from Sa-Sc, but not by very much.

\begin{figure}
\resizebox{\hsize}{!}{\rotatebox{-90}{\includegraphics{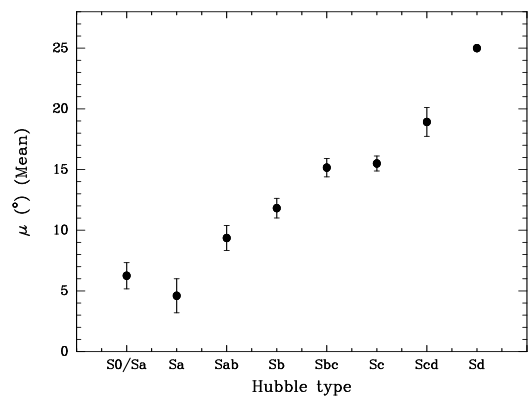}}}
\caption{Mean measured pitch angle plotted against different Hubble types in the RC3.}
\end{figure}

Fig.~7 presents the mean values versus the
Hubble types in the RC3, and it shows that, from the early to late Hubble
types, the mean values of the pitch angle increase, although the value for
S0/a's is larger than that for Sa's. In fact, it is difficult to distinguish
between S0/a and Sa.

Besides Hubble classification systems, Elmegreen \& Elmegreen (1982a, b; 1987)
introduced another classification system which is designed to emphasize
arm continuity, length and symmetry, and it is related to the presence
or strength of density waves. This system is not dependent on the galaxy Hubble
sequence and contains 12 distinct arm classes, corresponding to a
systematic change from the ragged and patchy arms in `flocculent' galaxies to
the two symmetric and continuous arms in `grand design' ones. Intermediate
arm classes show characteristics of both the `flocculent' and `grand design'
types. Based on their spiral arm classification system,
Elmegreen \& Elmegreen (1982a, b; 1987)
classified the spiral galaxies in the field, in binary systems, in
groups and in clusters, and suggested that bars tend to correlate with
spiral density waves, companions may influence (or generate) symmetric density
waves, grand design galaxies are physically larger than flocculent ones by
a factor of $\sim$ 1.5, and grand design galaxies are also preferentially
in dense groups.

\section{Conclusions}

In this paper, we investigate some statistical correlations about the thickness
of galactic disks and pitch angles of spiral arms. The main conclusions are:
(1). Early-type spirals, that are brighter on the average, are thicker;
(2). The axis ratio ($H_{z}/H_r$, here $H_{r}$ and $H_z$ are defined as two
times the radial scale length ($h_r$) and exponential scale height ($h$)) of
galactic disk
tends to be smaller along the Hubble sequence; (3). Except for a few
galaxies, early-type spiral galaxies have larger values of $H_{z}/H_r$;
(4). $H_{z}/H_r$ correlates strongly with galaxy color; (5). The
inclinations obtained by fitting the pattern of spiral
structure with a logarithmic spiral form are nearly the same as those obtained
by using the formulas of Aaronson et al. (1980); (6).
The mean measured pitch angles for different Hubble sequences
in the RC3 are derived, and the results show that the mean pitch angles
of Sa-Sc's are not larger than $15.5^{\circ}$, so that cotan$\mu\geq3.6$.
Thus the  WKB approximation can be satisfied at the mean pitch
angles from Sa-Sc's, but not by very much;
(7). From early to late Hubble types, the mean value of pitch angle
increases, despite some scattering.

Although the method, proposed by Peng (1988)
for deriving the thickness of
a face-on disc, is effective and
simple, it relies on a spiral structure theory
that predicts that a spiral arm has to end somewhere in the disk.
However, this theory might not be completely right. For example,
Zaritsky et al. (1993) has presented K-band (2.2 $\mu m$) images of M~51,
which reveal remarkable dynamical structure not visible in the
conventional optical observations, and show that the spiral arms
extend significantly
further towards the galaxy's center than previously
observed. In the optical images, the spiral arms begins
at a radius of about $30^{''}$ from the center. But,
the K-band residual image showed the spiral arms wind
through an additional $540^{\circ}$ beyond that
seen in the B-band images of the entire galaxy,
and end at about $10^{''}$ from the center of the galaxy. If
spiral galaxies whose spiral arms do
not stop anywhere in the disk exist, the parameters
$\rho_0$ and $r_0$ (Peng, 1988; Ma et al., 1997, 1998)
cannot be found.
In the K-band images,
which minimize the effect of dust and maximize sensitivity to
the dominant stellar population, we can derive our reliable values of
$\rho_0$ and $r_0$. As an example, we apply our
method to the image of M~51 in the K-band\footnote{The image of M~51 in
K-band is provided by Prof. Zaritsky.}
and derive the flatness
of disk ($H_z/D_0$), which is $0.010\pm 26.4\%$. Comparing this value
with Peng's (1988) ($H_z/D_0=0.013\pm 16.4\%$), we can
find that the value of the flatness based on the K-band image
is smaller. The reason is that, in the K-band image, the effect of
dust can be minimized and the values of $\rho_0$ and $r_0$ may
be reliably derived. We also emphasize that the images, which Ma et al.
(1998) used, are from the Digitized Palomar Sky Survey, in which many
images have burnt-out centers. There are some pictures that have
burnt-out centers in Ma et al. (1998), but we can change the
parameters of DISPLAY program to minimize the effect.

\begin{acknowledgements}
We are indebted to the anonymous referee for many critical
comments and helpful suggestions that have greatly
improved our paper, and for correcting our English.
We are grateful to Prof. Zaritsky for providing us
the image of M~51 in the K-band.
J. Ma gratefully acknowledges the support of K. C. Wong Education
Foundation, Hong Kong. This work is supported partly by the Chinese
National Science Foundation, No. 19603003.
\end{acknowledgements}


\begin{thebibliography}{}
\bibitem{} Aaronson M., Mould J., Huchra J., 1980, ApJ 237, 655
\bibitem{} Binggeli B., Sandage A., Tammann G. A., 1985, AJ 90, 1681
\bibitem{} Binggeli B., Sandage A., Tammann G. A., 1988, ARA\&A 26, 509
\bibitem{} Binney J., Tremaine S., 1987, Galactic Dynamics.
Princeton Univ. Press, Princeton
\bibitem{} Danver C.C., 1942, Ann. Obs. Lund. No. 10
\bibitem{} de Vaucouleurs G., 1956, Mem. Commonwealth Obs. (Mount Stromolo),
Vol. 3, No. 13
\bibitem{} de Vaucouleurs G., 1958, ApJ 128, 465
\bibitem{} de Vaucouleurs G., 1959, Handbuch der Physik 53, 275
\bibitem{} de Vaucouleurs G., de Vaucouleurs A., Corwin, H.G.Jr, 1976,
The Second Reference Catalogue of Bright Galaxies. University of Texas, Austin
\bibitem{} de Vaucouleurs G., de Vaucouleurs A., Corwin, H.G.Jr, et al.,
1991, The Third Reference Catalogue of
Bright Galaxies. Springer--Verlag, New York
(RC3)
\bibitem{} Dressler A., 1980, ApJS 42, 565
\bibitem{} de Grijs R., 1998, MNRAS 299, 595
\bibitem{} Elmegreen D. M., Elmegreen B. G., 1982a, MNRAS 201, 1021
\bibitem{} Elmegreen D. M., Elmegreen B. G., 1982b, MNRAS 201, 1035
\bibitem{} Elmegreen D. M., Elmegreen, B. G., 1987, ApJ 314, 3
\bibitem{} Freeman K. C., 1970, ApJ 160, 811
\bibitem{} Giovanelli R., Haynes M.P., Herter T., et al,
1997, AJ 113, 22
\bibitem{} Groot H., 1925, MNRAS LXXXV. 6, 525
\bibitem{} Grosb$\phi$l P.J., 1985, A\&ASS 60, 261
\bibitem{} Guthrie B.N.G., 1992, A\&ASS 93, 255
\bibitem{} Holmberg E., 1958, Medd. Lund. Obs. Ser.2, No. 136
\bibitem{} Hubble E.P., 1926, ApJ 64, 321
\bibitem{} Hubble E.P., 1936, The Realm of the Nebuloe. Yale Univ. Press, New Haven
\bibitem{} Ma J., Peng Qiu-he, Gu Qiu-sheng, 1997, ApJ 490, L51
\bibitem{} Ma J., Peng Qiu-he, Gu Qiu-sheng, 1998, A\&ASS 130, 449
\bibitem{} Kennicutt R.C., 1981, AJ 86, 1847
\bibitem{} Kennicutt R.C., Hodge P., 1982, ApJ 253, 101
\bibitem{} Morgan W.W., PASP 1958, 70, 364
\bibitem{} Morgan W.W., PASP 1959, 71, 394
\bibitem{} Lahav O., Naim A., Buta R.J., et al.,
1995, Sci 267, 859
\bibitem{} Naim A., Lahav O., Buta R.J., et al., 1995, MNRAS 274, 1107 (Sci 267, 859)
\bibitem{} Roberts M.S., Haynes M.P., 1994, ARA\&A 32, 115
\bibitem{} Peng Qiu-he, Li Xiao-qing, Huang Ke-liang, et al., 1979, Sci. in China XXII, 925
\bibitem{} Peng Qiu-he, 1988, A\&A 206, 18
\bibitem{} Sandage A.R., 1961, The Hubble Atlas of Galaxies.
Carnegie Institute of Washington, Washington
\bibitem{} Sandage A.R., Freeman K.C., Stokes N.R., 1970, ApJ 160, 831
\bibitem{} Sandage A.R., Tammann G.A., 1981, A Revised Shapley-Ames Catalog of
Galaxies. Carnegie Institute of Washington, Washington
\bibitem{} Sandage A.R., Tammann G.A., 1987, A Revised Shapley-Ames Catalog of
Bright Galaxies. Carnegie Institute of Washington, Washington
\bibitem{} Sandage A.R., Bedke J., 1993, Carnegie Atlas of Galaxies.
Carnegie Institute of Washington, Washington
\bibitem{} Shao Z.Y., Zhao.J.L., 1996, CAA 20, 273
\bibitem{} Shu C.G., Zhao J.L., Tian K.P., 1995, Acta Astrophy. Sin. 15, 110
\bibitem{} Tully R.B., Fisher J.R., 1977, A\&A 54, 661
\bibitem{} van der Kruit P.C., Searle L., 1981a, A\&A 95, 105
\bibitem{} van der Kruit P.C., Searle L., 1981b, A\&A 95, 116
\bibitem{} van der Kruit P.C., Searle L., 1982a, A\&A 110, 61
\bibitem{} van der Kruit P.C., Searle, L., 1982b, A\&A 110, 79
\bibitem{} ven den Bergh S., 1960a, ApJ 131, 215
\bibitem{} ven den Bergh S., 1960b, ApJ 131, 558
\bibitem{} ven den Bergh S., 1976, ApJ 206, 883
\bibitem{} von der Pahlen E., 1911, Astron. Nachr. 188, 249
\bibitem{} Wilkerson M.S., 1980, ApJ 241, L115
\bibitem{} Zaritsky D., Rix H.W., Rieke M., 1993, Nat 364, 313
\bibitem{} Zhao J.L., Shao Z.Y., 1994, A\&A 289, 89
\end{thebibliography}
\end{document}